# Chandojnanam: A Sanskrit Meter Identification and Utilization System

**Hrishikesh Terdalkar** **Arnab Bhattacharya**
hrishirt@cse.iitk.ac.in arnabb@cse.iitk.ac.in
Department of Computer Science and Engineering,
Indian Institute of Technology Kanpur,
India

## Abstract

We present Chandojñānam, a web-based Sanskrit meter (chanda) identification and utilization system. In addition to the core functionality of identifying meters, it sports a friendly user interface to display the scansion, which is a graphical representation of the metrical pattern. The system supports identification of meters from uploaded images by using optical character recognition (OCR) engines in the backend. It is also able to process entire text files at a time. The text can be processed in two modes, either by treating it as a list of individual lines, or as a collection of verses. When a line or a verse does not correspond exactly to a known meter, Chandojñānam is capable of finding fuzzy (i.e., approximate and close) matches based on sequence matching. This opens up the scope of a meter based correction of erroneous digital corpora. The system is available for use at `https://sanskrit.iitk.ac.in/jnanasangraha/chanda/`, and the source code in the form of a Python library is made available at `https://github.com/hrishikeshrt/chanda/`.

## 1 Introduction

Majority of the Sanskrit literature is in the form of poetry that adheres to the rules of Sanskrit prosody or Chandaḥśāstra, which is the study of Sanskrit *meters*, known as chandas. The purpose of chanda is primarily to add rhythm to the text so that it is easier to memorize. Additionally, it also helps in preserving the correctness to some extent.

In this paper, we introduce a web-based system, Chandojñānam, to detect the chanda in Sanskrit text. The text can be either a single line or a verse or even an image. With increasing efforts towards digitization of Sanskrit literature, our system serves as the go-to portal to detect, study and analyze vṛttas in Sanskrit text.

The digitization of Sanskrit text is achieved primarily through two methods: (1) manual data entry and (2) scanning of documents followed by optical character recognition (OCR). The former suffers from human error while the latter is prone to inaccuracies due to automated processing. Further, with the rise of social media, blog sites and Unicode, there is a large replication of Sanskrit text on the Internet with little quality control, thereby further increasing errors in the text.

We argue that a non-trivial portion of the errors introduced in texts through various sources can be detected by the process of meter identification. The Chandojñānam system exhibits tolerance towards erroneous texts and is able to locate the errors as well as make suggestions for fixing them.

### 1.1 Motivation

The motivation behind Chandojñānam can be better understood with the following scenarios.

- A Sanskrit enthusiast wants to identify the meter of a verse from a PDF file with Sanskrit text. She however, lacks the capability to effectively type the Sanskrit text, and prefers to upload a screenshot of the specified verse to identify the meter. With Chandojñānam, she is able to perform meter identification directly from the image.

- A teacher of Chandaḥśāstra wants to explain the rules of Sanskrit prosody to her students using several examples. She does not want to spend precious time on writing down all the intermediate steps in the identification of a meter. Instead, she wants a system that will do this job for her. Chandojñānam fits the bill nicely.

- A budding poet is trying to compose Sanskrit poetry adhering to a specific meter. However, she is not an expert, and may have made some errors. She wants to locate these errors so that she can correct it. Chandojñānam lets her locate these errors, and also provides suggestions for correction.

- A Sanskrit researcher wants to analyse a large Sanskrit text file and obtain metrical statistics of the entire corpus. Chandojñānam allows her to upload the text file and quickly obtain the required statistics to aid her in her research.

These examples highlight the utility of Chandojñānam, in addition to just satisfying one's curiosity about Sanskrit meters.

### 1.2 Background

The classification of syllables into laghu (*short*) and guru (*long*) forms the core concept of Chandaḥśāstra Deo (2007). The classification is related to the amount of time it takes to pronounce a specific syllable; in particular, the short syllables are termed laghu and the long syllables guru. Specific sequences or combinations of laghu and guru letters result in a particular rhythm or chanda.

A verse (śloka) is composed of four parts, each known as a pāda. Every meter has a constraint on the sequence of syllables that should be followed in a pāda. For example, the meter Pañcacāmara is defined by each pāda having 16 syllables with the following laghu-guru sequence of syllables: `LGLGLGLGLGLGLGLG`[1]. The perfect alternation of laghu and guru syllables is a reason for its energetic tempo, as can be experienced in the hymns such as Śivatāṇḍavastotram[2] or Narmadāṣṭakam[3]. In a similar manner, numerous Sanskrit meters have been defined in the texts on Sanskrit prosody such as Vṛttaratnākara.

The unique sequence of laghu-guru markers used to identify a meter is hereon referred to as *lg-signature* of that meter. The term, in reference to an arbitrary Sanskrit line, is used to refer to its decomposition in the laghu-guru sequence. Chandaḥśāstra, for brevity and ease of remembering, identifies all laghu-guru sequences of length 3 with a unique letter. This unique 3-length sequence is known as a Gaṇa. The possible number of gaṇas is $2^3 = 8$. More details about the terminology and rules of Chandaḥśāstra can be found at `https://sanskrit.iitk.ac.in/jnanasangraha/chanda/help`. Several previous works (Deo, 2007; Mishra, 2007; Melnad et al., 2013) also discuss the theory of Chandaḥśāstra in detail.

### 1.3 Related Work

There have been several efforts in the area of automatically identifying meter from Sanskrit text. Some of these tools (Mishra, 2007; Melnad et al., 2013) were only presented as web interfaces, which are no longer functional.

More recent works (Rajagopalan, 2020; Neill, 2022) provide both a web interface and a software library. However, the web interfaces provided by Rajagopalan (2020) and Neill (2022) both assume that a single verse will be provided as an input, limiting its usability to quickly check meters for a set of verses. Thus, even for a text consisting of a small number of verses, the input has to be provided as many times as there are number of verses. Neill (2022) attempts to address this issue by allowing upload of text files. However, it still lacks a display in case one wants to visualize meters for multiple verses. More significantly, identifying the meter of a single pāda or a partial verse is not possible at all, as any text entered is assumed to consist of exactly four pādas.

[1] We use letters `L` and `G` to denote laghu and guru syllables respectively.
[2] `https://shlokam.org/shivatandavastotram/`
[3] `https://shlokam.org/narmadashtakam/`

| Features | | Mishra (2007) | Melnad et al. (2013) | Rajagopalan (2020) | Neill (2022) | Chandojñānam |
|---|---|---|---|---|---|---|
| Availability | Web Interface | ✓[4] | ✓[5] | ✓ | ✓ | ✓ |
| | Software Library | | | ✓ | ✓ | ✓ |
| Input | Text | ✓ | ✓ | ✓ | ✓ | ✓ |
| | Arbitrary Lines | | | | | ✓ |
| | Multiple Verses | | | | | ✓ |
| | Textfile Upload | | | | ✓ | ✓ |
| | Image Upload | | | | | ✓ |
| Functionality | Meter Identification | ✓ | ✓ | ✓ | ✓ | ✓ |
| | Error Tolerance | | | ✓ | ✓ | ✓ |
| | Fuzzy Matching | | | ✓ | | ✓ |

Table 1: Feature comparison of extant meter identification systems

Error tolerance, i.e., the capacity to identify the meter of a verse in the presence of errors, is an important requirement for a meter identification system. To be useful for error correction, a system should be further able to detect exact locations of error and suggest corrections. Neill (2022) uses a scoring system resembling majority rule on the pādas of a verse to identify its meter. However, it assumes pāda matching to be exact. In other words, even if there is a single error in a pāda, the system reports that pāda as a mismatch. As a result, when there are errors in three pādas, due to the majority of pādas being mismatched, the overall meter for the verse is reported as '*unknown*'. Further, it lacks any suggestion mechanism. Barring the admirable work by Rajagopalan (2020), none of the other tools attempts to perform *fuzzy matching* (i.e., *approximate matching*) to handle erroneous text. The chanda suggested by Rajagopalan (2020) is, however, the one that is perceived as the best by its fuzzy matching algorithm, which may not always be the actual chanda of the verse. Therefore, providing a top-$k$ ranked list of matches is more useful.

Further, none of the existing tools has a functionality to upload images for meter identification.

Chandojñānam attempts to overcome the shortcomings of the previous tools by providing a comprehensive set of user-friendly features. There is a special focus on *fuzzy matching* (explained in §2.4.2) and the utility of Chandaḥśāstra for correction of digital corpora.

Table 1 presents a feature matrix comparing the other works with Chandojñānam.

## 1.4 Contributions

We present, Chandojñānam, a Sanskrit meter identification and utilization system, which in addition to identifying a meter, also aims to catch errors in the text and suggest corrections. The aim of the tool is to make this process easy for a non-programmer. The salient features of Chandojñānam can be summarized as follows:

- The tool is available as a web-application that can be used from any standard browser and requires no other installation from the user. The library is also made available for the benefit of programmers.
- There are three prominent input options: (1) plain text, (2) images (screenshots) and (3) text files.
- The input can be in any of the standard transliteration scheme that can be detected by the

[4] `http://sanskrit.sai.uni-heidelberg.de/Chanda/HTML/` is no longer functional.
[5] `https://sanskritlibrary.org:8080/MeterIdentification/` is no longer functional.

`indic-transliteration` (Sanskrit-programmers (2021)) library. This applies to all three input methods.

- The image files can be processed using one of the two OCR engines, namely, Google OCR and Tesseract OCR, and the detected text can be further edited.
- The lines from input are detected based on several standard line-end markers, such as '`\n`', '।', '॥' and '`.`'
- Meter identification can be performed on the line (pāda) level. The input is treated as a set of lines. Therefore, the input can be any arbitrary set of pādas.
- Meter identification can also be performed on a verse (śloka) level. In this case, the system treats the lines as being parts of a verse. Cumulative cost of each line is minimized to identify the meter of the verse. Multiple verses can be provided.
- For erroneous inputs, the tool provides a robust fuzzy matching support using edit distance comparison, which helps in identifying and highlighting the places where an error might be present. Additionally, there is a suggestion module to help the user understand what changes can be made to the input.
- Informative display shows the steps involved in the meter identification, which is aimed to help learners of Chandaḥśāstra.
- Results can be downloaded in the JSON format.

The Chandojñānam system is a available online at `https://sanskrit.iitk.ac.in/jnanasangraha/chanda/`. The source code can be found at `https://github.com/hrishikeshrt/chanda/`.

## 2 The Chandojñānam System

In this section, we discuss the inner workings of the Chandojñānam system. Figure 1 illustrates the overall workflow of the system. Initially, definitions of Sanskrit meters are read into the system and stored in the form of a dictionary, referred to hereon as the *metrical database*. A user may specify the input in any of the three specified formats, namely, plain text, image containing text, and text file. For image, the text is extracted using OCR systems. After the text is extracted, the transliteration scheme is detected, and converted to the internal transliteration scheme, which is 'Devanagari'. From the text, lines are detected which, in turn, are split into syllables to obtain the *lg-signature* of the lines. The metrical database is then queried using the meter detection algorithm and the matches are presented to the user along with useful information. In case of erroneous inputs, $k = 10$ closest matches are also shown, along with the suggestions for corrections.

### 2.1 Chanda Definitions

There are two main types of meters, Varṇavṛtta and Mātrāvṛtta (Melnad et al., 2013). Currently, the system only deals with Varṇavṛtta. In future, we will enable the system to handle Mātrāvṛtta as well. Varṇavṛtta can be further divided into three categories, namely, Samavṛtta, Ardhasamavṛtta and Viṣamavṛtta. This categorization is performed based on the symmetry or lack thereof of the metrical pattern exhibited by the four pādas of a śloka adhering to the meter. All four pādas following the same pattern is termed as Samavṛtta, odd and even pādas following a different pattern is termed as Ardhasamavṛtta, and all four pādas following a different pattern corresponds to Viṣamavṛtta.

Chanda definitions are specified in the tabular format as illustrated in Figure 2. The legend is described in Table 2. It can be noted that the value corresponding to the column pāda denotes the index of pāda in the meter described in that row. This way of specifying definitions results in a uniform treatment of samavṛtta, ardhasamavṛtta and viṣamavṛtta. Additionally, a regex pattern (*regular expression*) definition can also be specified, where the metrical restriction only applies to a part of the pāda. The most commonly used meter in Sanskrit, namely, anuṣṭubh, is a prime example of such regex patterns. Specifically, the requirements for anuṣṭubh chanda are:

- Every pāda must contain exactly 8 syllables.
- The 5th syllable of every pāda must be a laghu syllable.
- The 6th syllable of every pāda must be a guru syllable.
- The 7th syllable of the even pādas must be a laghu syllable.
- There is no other restriction on the other syllables, i.e., they can be either laghu or guru.

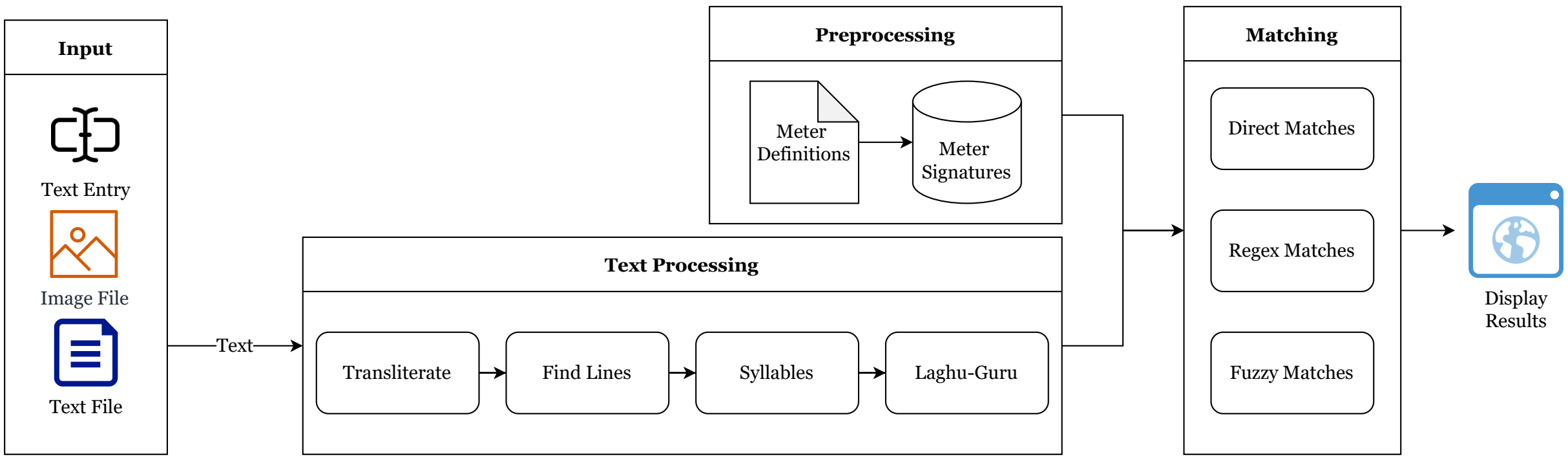

Figure 1: Workflow of the Chandojñānam system

| वृत्त | पाद | गण | लक्षण | अक्षरसङ्ख्या | मात्रा | यति |
|---|---|---|---|---|---|---|
| शार्दूलविक्रीडित | | मसजसततग | गगगललगलगलललगगगलगगलग | 19 | 30 | 12,7 |
| शालिनी | | मततगग | गगगगगलगगलगग | 11 | 20 | 4,7 |
| अपरवक्त्र | 1 | ननरलग | ललललललगलगलग | 11 | 14 | |
| अपरवक्त्र | 2 | नजजर | ललललगललगलगलग | 12 | 16 | |
| सौरभ | 1 | सजसल | ललगलगलललगल | 10 | 13 | |
| सौरभ | 2 | नसजग | लललललगलगलग | 10 | 13 | |
| सौरभ | 3 | रनभग | गलगलललगललग | 10 | 14 | |
| सौरभ | 4 | सजसजग | ललगलगलललगलगलग | 13 | 18 | |
| अनुष्टुभ् | 1 | ----लग-- | ----लग-- | 8 | | |
| अनुष्टुभ् | 2 | ----लगल- | ----लगल- | 8 | | |

Figure 2: Generic Chanda definition format

Therefore, the regex patterns `[LG][LG][LG][LG]LG[LG][LG]` and `[LG][LG][LG][LG]LGL[LG]` correspond to the *lg-signatures* for the odd numbered and even numbered pādas of Anuṣṭubh respectively.

Internally, the meter database is stored in the form of dictionaries in two ways: signature of individual pādas is stored as `CHANDA_SINGLE`, while signature of consecutive pādas is stored as `CHANDA_MULTIPLE`. It may often happen that two pādas of the input verse are given as a single line without any kind of line marker. The second dictionary acts as a fail-safe when the lines couldn't be split in a proper manner due to lack of punctuation or line-breaks.

| Column | Requirement | Description |
|---|---|---|
| Vṛtta | required | Name of the meter described in the row |
| Pāda | required | Index of pāda in the corresponding vṛtta; the possible values are <blank>, 1, 2, 3 and 4. |
| Lakṣaṇa | required | *lg-signature* of the meter |
| Gaṇa | optional | Signature of the meter in the compressed (trika) notation |
| Akṣarasaṃkhyā | optional | Number of letters in the pāda |
| Mātrā | optional | Number of mātrās in the pāda |
| Yati | optional | Indices corresponding to yati |

Table 2: Chanda Definitions specification format

For example, for the meter Bhujaṅgaprayāta, which has a signature यययय[6], corresponding to the *lg-signature* of `LGGLGGLGGLGG`, two independent entries are maintained.

```
CHANDA_SINGLE = {
    'LGGLGGLGGLGG': ['Bhujaṅgaprayāta'],
    '[LG][LG][LG][LG]LG[LG][LG]': ['Anuṣṭubh (Pāda 1)'],
    '[LG][LG][LG][LG]LGL[LG]': ['Anuṣṭubh (Pāda 2)']
}
CHANDA_MULTIPLE = {
    'LGGLGGLGGLGGLGGLGGLGGLGG': ['Bhujaṅgaprayāta (Pāda 1-2)'],
    '[LG][LG][LG][LG]LG[LG][LG][LG][LG][LG][LG]LGL[LG]': ['Anuṣṭubh (Pāda 1-2)']
}
```

Currently, the system has data for more than 200 meters. This number is less as compared to the database used by Rajagopalan (2020); however, the detailed discussion of why more data need not always mean better is also provided by Rajagopalan (2020). The primary reason for not importing a large number of meters is that the lower number results in less false positives when dealing with erroneous input. More meters may be included in the system in future based on user feedback.

### 2.2 Input

The input for meter identification is Sanskrit text, and it can be provided to our system in three ways. The simplest form of input is a direct text entry, wherein, a user may type or copy-paste Sanskrit text in any valid transliteration scheme into a text input box.

Perhaps a more useful option, which is a novelty of our tool, is the ability to process images containing text. Two well-known OCR engines, Google Drive OCR[7] and Tesseract OCR[8] (Kay, 2007) are supported. Google Drive OCR functions by making use of Google's Drive API v3[9] to upload files to Google Drive, while Tesseract OCR v5 is a neural network (LSTM) based OCR engine. Python libraries `google-drive-ocr`[10] (Terdalkar, 2022) and `pytesseract`[11] (Lee, 2022) are utilized respectively to access the OCR systems. Google Drive OCR is generally more accurate than Tesseract OCR but slower due to the network latency imposed by upload of files. The output of either of the OCR systems, i.e., the recognized text, is treated as if directly entered by the user. The output is not always accurate, especially if the image contains noisy text. Therefore, the system also lets user edit the optically recognized text and re-submit it to the system. Figure 3 shows the interface to upload images for meter identification.

[6] Gaṇa य corresponds to the *lg-sequence* `LGG`.
[7] `https://support.google.com/drive/answer/176692?hl=en`
[8] `https://github.com/tesseract-ocr/`
[9] `https://developers.google.com/drive/api/v3/reference`
[10] `https://github.com/hrishikeshrt/google_drive_ocr`
[11] `https://github.com/madmaze/pytesseract`

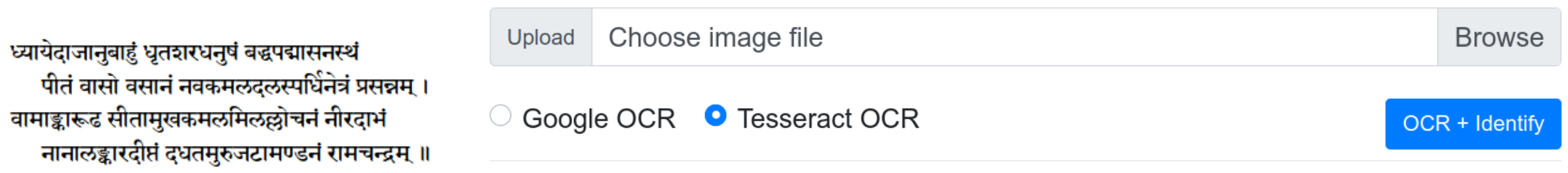

Figure 3: Upload a screenshot of a verse to Chandojñānam for meter identification

For bulk processing, the option to upload a text file is also available. After reading the uploaded file, the text is again treated as if directly entered by the user.

The identical treatment of all input allows the same meter identification pipeline to follow. For example, as the transliteration module is triggered after the processing of input text, all three input methods have full support for input in any valid transliteration scheme.

### 2.3 Text Processing

Once the text input is obtained and cleaned, it passes through four primary processing steps.

1. **Transliteration:** The problem of transliteration of Devanagari text can be considered to be a solved one due the presence of robust tools such as `indic-transliteration`[12] (Sanskrit-programmers, 2021) and `Aksharamukha`[13] (Rajan, 2018). We rely on `indic-transliteration` for detection of input scheme; thus, any transliteration scheme detectable by `indic-transliteration` is also supported by our system. The internal transliteration scheme is set to 'Devanagari', which is a convenience choice as we believe that it makes the code easier to follow.

2. Pāda **Split:** The contiguous text is now split into lines (pādas) based on several standard line-end markers, such as '\n', '।', '॥' and '.'. Chandojñānam uses pāda as a unit for meter identification.

3. **Syllabification**: The process of splitting text into syllables is fairly straightforward for majority of Indian languages due to phonetic consistency of the alphabet. The Devanagari Unicode has special vowel markers for all vowels except 'अ' (a). The presence of vowels with consonants, therefore, can be easily identified with these markers, and syllables can be separated. The absence of vowel marker for the vowel 'अ' (a) can be treated by noting that for a joint consonant, Unicode always uses the halanta marker '◌्'. So, two consecutive consonant characters indicate the presence of a vowel 'अ' with the first consonant. For example, the Unicode string "भारत" consists of 4 Unicode characters: 'भ', '◌ा', 'र' and 'त'. However, it corresponds to 6 Devanagari letters (varṇa), namely, 'भ्', 'आ', 'र्', 'अ', 'त्' and 'अ'. Every vowel signifies an end of a syllable, resulting in three syllables, namely, 'भा', 'र' and 'त'. Thus, in a single scan of the string, syllabification can be performed. We use the `sanskrit-text`[14] Python library to perform this task.

4. Laghu-Guru **Marks**: The syllabification process is a prerequisite for obtaining the *lg-signature* of a Sanskrit line. Standard rules of Sanskrit prosody described in numerous articles (Deo, 2007; Melnad et al., 2013; Rajagopalan, 2020) are followed to mark each syllable as laghu or guru. A general rule of Piṅgala's Chandaḥśāstra states that the last syllable of a pāda should be treated as a guru. However, there exist meters in the Chandaḥśāstra whose signatures contain the last syllable as laghu (Pādānta Laghu). Therefore, we compute the *lg-signature* without forcing the last letter to be guru.

[12] `https://pypi.org/project/indic-transliteration/`
[13] `https://aksharamukha.appspot.com/`
[14] `https://pypi.org/project/sanskrit-text/`

**Algorithm 1:** Meter Identification

**Data:** Metrical Database ($MD$)
**Input:** *lg-signatures* corresponding to each 'line' in the input ($T = \{lg_1, lg_2, \ldots, lg_n\}$)
**Output:** Result set containing exact or fuzzy matches

```
forall lg ∈ T do
    SM_1 = FindDirectMatch(lg, 'CHANDA_SINGLE')
    SM_2 = FindDirectMatch(lg, 'CHANDA_MULTIPLE')
    RM = FindRegexMatch(lg, 'CHANDA_SINGLE' + 'CHANDA_MULTIPLE')
    DM = SM_1 + SM_2 + RM
    FM = ϕ
    if DM = ϕ then
        FM = FindFuzzyMatch(lg)
    end
    return DM + FM
end
```

**Algorithm 2:** Direct Matching

**Input:** *lg-signature*
**Output:** Result set containing exact matches

```
Function FindDirectMatch(lg, 'MD')
    M_1 = Query(lg, 'MD')          // dictionary lookup
    M_2 = ϕ
    if M_1 = ϕ then                // if no match found
        if the last letter of lg is laghu then
            lg_1 = replace last letter of lg with guru
            M_2 = Query(lg_1, 'MD')
        end
    end
    return M_1 + M_2
```

This completes the text processing which makes the text ready for the next stage, which is meter identification.

### 2.4 Meter Identification Algorithm

The *meter identification* problem, in its simplest form, is a dictionary lookup on *lg-signatures* of meters. In practice, however, there can be several additional considerations. The algorithm used by our system is described in Algorithm 1. The algorithm consists of a direct dictionary lookup, regular-expression based lookup and fuzzy matching. These are explained in the subsequent sections.

#### 2.4.1 Direct Matching

Finding a direct match involves trying to find the match in the both the dictionaries in the metrical database that store the meter signatures. Further, for meters with regex specification, regular-expression-based matching is performed.

As mentioned in §2.3, we compute the *lg-signature* without enforcing the last syllable to be guru. If the last syllable was laghu and no direct match was found, we attempt to find a direct match by treating the last syllable as guru. Algorithm 2 describes this process. The function `Query` corresponds to a simple dictionary lookup and returns a set of matches.

In theory, this should be sufficient. However, in practice, one often encounters erroneous text. This is expected since a large amount of Sanskrit text available digitally is a result of either manual entry or post-scanning OCR followed by manual correction. As a result, due to human errors or OCR inaccuracies, the following errors may be found in the texts available online.

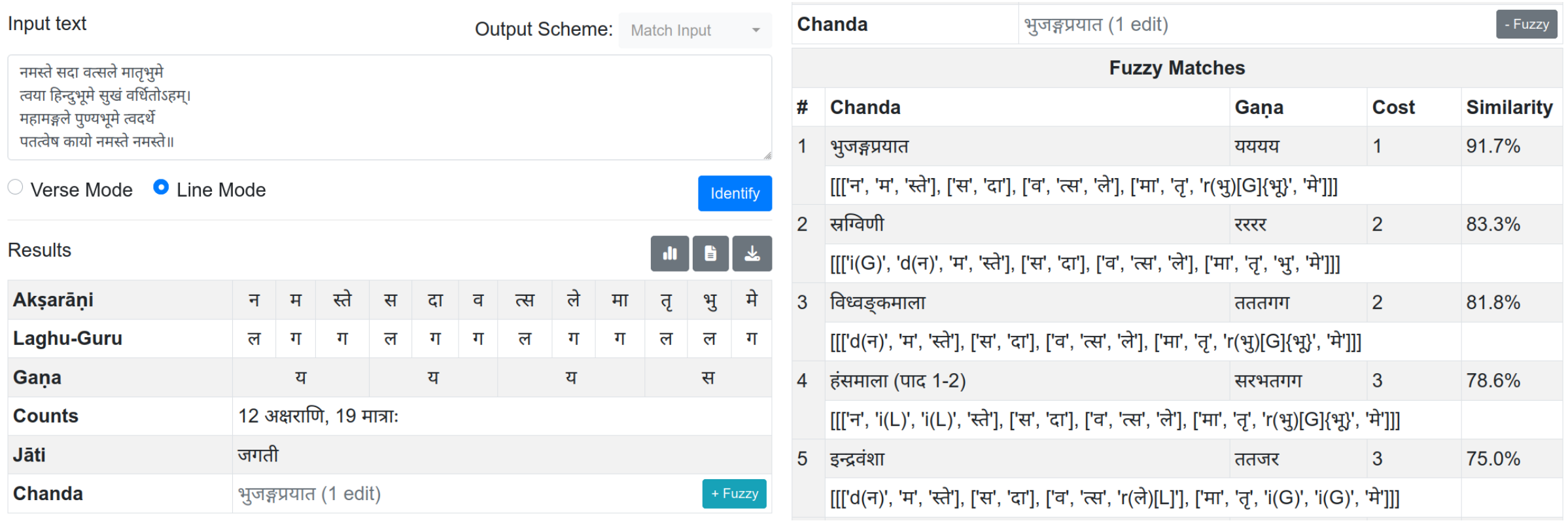

Figure 4: Meter identification with fuzzy matching and suggestions

- Characters may be misspelt, e.g., रु (ru) as रू (rū)
- Characters may be missing, e.g., वर्गै (vargai) as वगै (vagai)
- Characters may be misidentified, e.g., ऋ (ṛ) as क्र (kra)
- Characters may get split, e.g., ख (kha) as रव (rava)

Most of these errors may also affect the metrical pattern of a line and, therefore, meter identification is a useful tool to identify them. It should be mentioned here that not all errors lead to metrical mismatch. For example, an error such as the confusion between letter pairs such as व, ब or म, स would not affect the meter pattern and, thus, cannot be captured by only meter identification.

### 2.4.2 Fuzzy Matching

For non-exact matches, we model the problem as that of finding the *nearest matching string* for the *lg-signature* of the text. In particular, we compute the Levenshtein edit distance of the observed pattern with all the known patterns. We then find the similarity by first normalizing the edit distance using the length of the target match, and then subtracting it from 1.

$$\text{Similarity} = 1 - \frac{\text{Levenshtein distance}}{\text{length of target match}}$$

We present the topmost $k$ matches as the possible fuzzy matches (currently, $k = 10$). Algorithm 3 and Algorithm 4 describes the core functions pertaining to fuzzy matching. The *Similarity* column in the interface (Figure 4) shows the similarity instead of the edit distance.

For finding the edit-distance and edit-operations, we use the `python-Levenshtein`[15] library (Haapala, 2014). The `AddEditOpsMarkers` function (line 5) formats the original list of syllables by adding suggestions based on the edit operations. The suggestions shown contain the characters from the original string, along with the suggested places where a change might be required. Suggested changes for each meter correspond to the changes needed to transform the given line into that meter.

The suggestions are identified by three letters `i`, `d` and `r`, indicating *insert*, *delete* and *replace* operations respectively. They are used in the following manner.

[15] `https://pypi.org/project/python-Levenshtein/`

**Algorithm 3:** Fuzzy Matching

**Input:** *lg-signature* and $k$
**Output:** Result set containing top-$k$ fuzzy matches

**Function** `FindFuzzyMatch(`*lg, 'MD', k*`)`
    results = $\phi$
    **forall** *chanda in 'MD'* **do**
        cost, suggestion = `Transform(`*lg, lg-signature of chanda*`)`
        **if** *suggestion* **then**
            results += (chanda, cost, suggestion)
        **end**
    **end**
    **return** top-$k$ results with lowest cost

**Algorithm 4:** Transform String Sequences

**Input:** String sequences $seq_1$, $seq_2$
**Output:** Suggested edit operations required to transform $seq_1$ into $seq_2$ and the cost of transformation

**Function** `Transform(`$seq_1$, $seq_2$`)`
    edit_ops = GetLevenshteinEditOps($seq_1$, $seq_2$) `// using python-Levenshtein`
    weights = {'replace': 1, 'delete': 1, 'insert': 1}
    cost = $\sum_{op \in \text{edit_ops}}$ weights[op]
    suggestion = AddEditOpsMarkers($seq_1$, edit_ops) `// add suggestions`
    **return** cost, suggestion

- `i(L/G)`:
  This notation indicates the insertion of a new syllable. `i(L)` (respectively, `i(G)`) indicates that a laghu (respectively, guru) syllable needs to be inserted.
- `r(letter)[L/G]{suggestion}`:
  This notation indicates that the syllable needs to be replaced by a laghu or a guru syllable (depending upon the marker). In some of the cases, a suggested change of syllable may be included. A simple heuristic that is followed replaces the laghu syllables corresponding to the laghu vowels 'इ', 'उ' and 'ऋ' by their guru vowel counterparts, namely, 'ई', 'ऊ' and 'ॠ', respectively, and vice versa.
- `d(letter)`:
  This notation indicates that the syllable needs to be deleted.

The process of fuzzy matching can be better understood with the help of an example. Consider the first line of the example depicted in Figure 4.

- Line: नमस्ते सदा वत्सले मातृभुमे
- LG-Signature: `LGGLGGLGGLLG`
- Nearest Match: `LGGLGGLGGLGG` (cost 1) (Bhujaṅgaprayāta)
- Suggestion: [['न', 'म', 'स्ते'], ['स', 'दा'], ['व', 'त्स', 'ले'], ['मा', 'तृ', 'r(भु)[G]{भू}', 'मे']]

The exact match for the line is not found in the system and, therefore, we find the edit-distance of this *lg-signature* with every *lg-signature* from the database. It is found that the *lg-signature* `LGGLGGLGGLGG` of Bhujaṅgaprayāta meter is the closest and is only 1 edit-distance away. The required change is to replace the 11th letter from `L` (laghu) to `G` (guru). It is also suggested to replace the laghu letter भु into a guru letter भू.

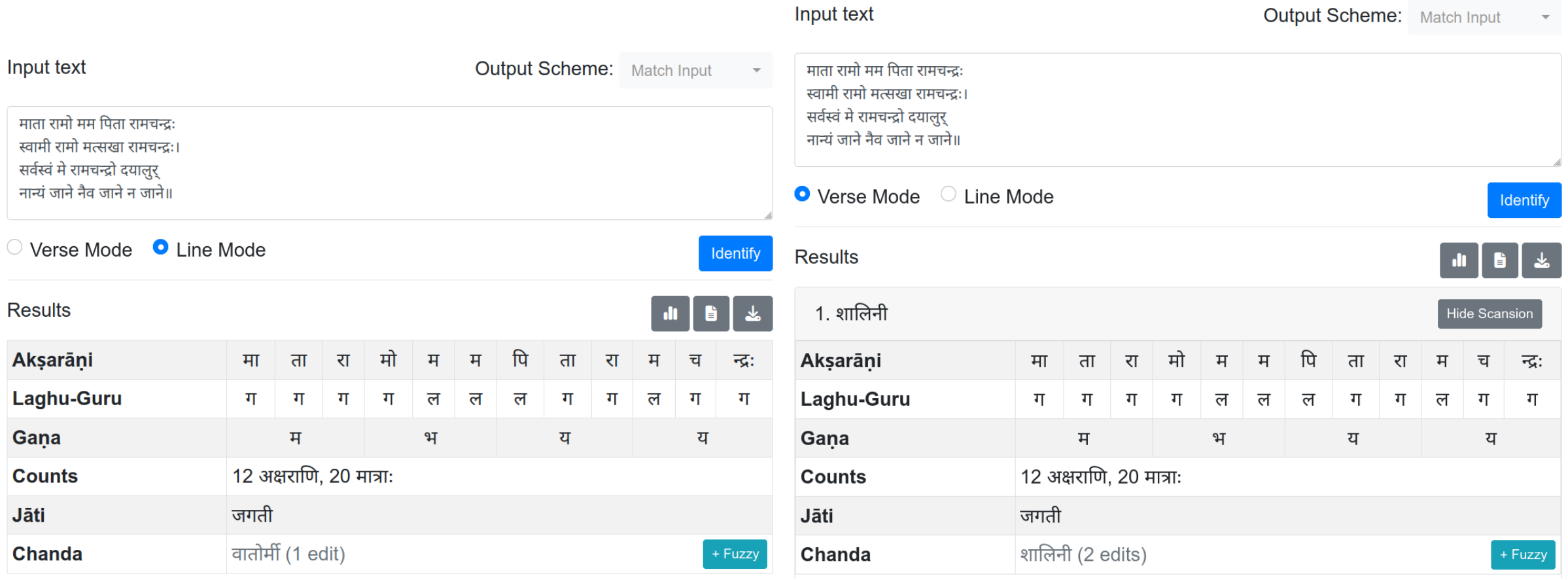

Figure 5: Meter identification from a verse in (a) *Line mode* and (b) *Verse mode*

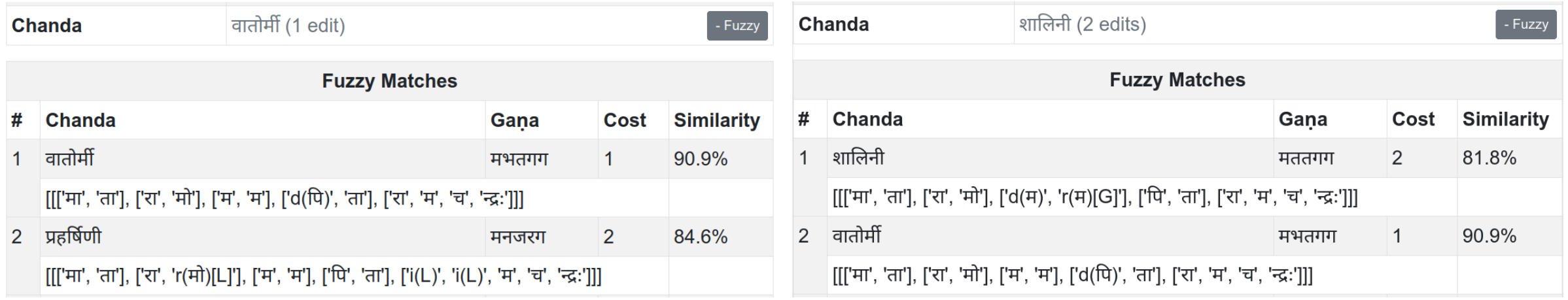

Figure 6: Fuzzy matches in (a) *Line mode* and (b) *Verse mode*

### 2.4.3 Verse Processing

Meter identification can be performed by treating the input either as a set of arbitrary lines (*Line mode*) or as a collection of verses (*Verse mode*). The *Line mode* is useful for checking meter of a single line or a set of lines. The treatment of the text as a verse differs insofar as it attempts to minimize the cumulative cost of the meter matches over each line.

Consider a sample verse adhering to the meter Śālinī, albeit with a deliberate small error in the first pāda. Figure 5 highlights the difference between the two modes of meter identification. The correct word मत्पिता has been replaced by two words मम पिता. As a result, the first pāda does not find an exact match. Further, the closest match based on edit distance is found to be the meter Vātormī with an edit distance of 1. Therefore, when in the *Line mode*, the suggestion is made as Vātormī. However, rest of the three pādas are exact matches for Śālinī. Therefore, the cumulative cost of Śālinī over the entire śloka is the lowest (here, it is 2). Thus, in the *Verse mode*, the identified meter for the verse is Śālinī. This highlights the utility of the *Verse mode*.

In the list of fuzzy matches shown, the meter identified for the verse is shown first. Figure 6 shows that the meter Śālinī is shown for the first pāda despite having a higher edit cost.

## 2.5 Output

A blackbox that only identifies meter would not be as interesting as a system that also explains ('teaches') the steps. Therefore, in addition to the meter name, scansion details such as *lg-signature*, gaṇa-signature, count of letters and count of mātrās are also displayed. Additional information displayed in the case of fuzzy matches is already discussed in §2.4.2. Entire result is displayed in a neat tabular format (as displayed in Figure 4) which makes the system useful for learners of Chandaḥśāstra.

Cumulative statistics such as number of lines identified, number of lines not identified, frequency distribution of exact and fuzzy matches etc. can also be viewed. Additionally, the data can also be downloaded. The data download supports two formats, a compact format for quick visual inspection, and a detailed JSON format for further computational processing.

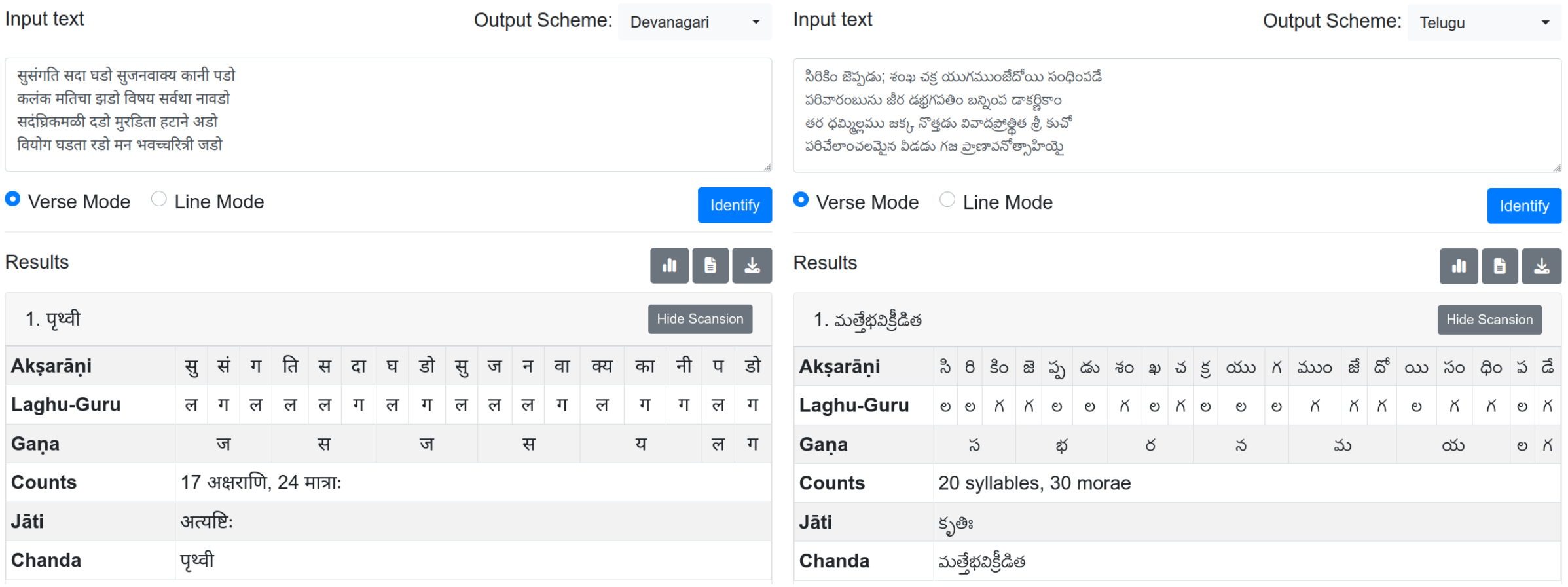

Figure 7: Meter identification from other Indian languages, e.g., (a) Marathi (b) Telugu

## 2.6 Utility

We can identify several use-cases for Chandojñānam, some of which have already been discussed in §1.1:

- The primary use-case, naturally, is the ability to identify meters from Sanskrit text.
- Ability to upload of images and identification of meters from it is another important scenario.
- Identification of errors from the Sanskrit text which has been created through the process of OCR, but not manually corrected yet, serves as another prominent use-case.
- Additionally, the system can be used to obtain metrical statistics from large texts automatically. These statistics may also help a user identify errors, if any, at a quick glance. For example, in the case of texts such as Meghadūta, which follows a single meter, namely, Mandākrāntā, any anomaly will be quickly spotted by appearance of a different meter in the statistics.
- A user may be interested in creating poetry in Sanskrit, and may require assistance in quickly checking the metrical consistency of her creation.
- Learners of Chandaḥśāstra may want to try out several examples to understand the process of meter identification in-depth.
- Several Indian languages, e.g., Marathi, Telugu, etc., exhibit rules of prosody similar to Sanskrit. Due to comprehensive transliteration support provided by the `indic-transliteration` library, text written in scripts other than Devanagari can be also used in the same manner. Figure 7 illustrates the usage of Chandojñānam for meter identification from Marathi and Telugu. Here, we are assuming that the same metrical database is used. However, there may be language-specific differences in the rules of prosody as well as variety of meters. Hence, the multilingual support is still primitive.

# 3 Evaluation for Error Correction

One of the primary goals of Chandojñānam is to facilitate error detection from the Sanskrit text obtained from various sources. Consequently, we evaluate the ability of the tool to correctly identify the meters from erroneous text.

### 3.1 Corpus

Evaluation of a meter identification system requires tagged metrical data. As mentioned by Rajagopalan (2020), Meghadūta composed by Kālidāsa (Kale, 2011), being entirely in Mandākrāntā meter, is an ideal corpus for evaluation. The same text, available from various sources, can also differ greatly in terms of encoding, character-by-character comparison and errors present. We use three online sources, namely, Wikisource[16], sanskritdocuments.org[17] and GRETIL[18] to obtain three versions of Meghadūta.

In addition to Meghadūta, we also use texts with more metrical variety. We choose three texts from Wikisource, namely, Śāntavilāsa[19], Śrīrāmarakṣāstotra[20] and Rājendrakarṇapūra[21]. We manually tag meters for each verse from these texts. Together, the 4 datasets contain 1038 verses, exhibiting 17 distinct meters.

Further, to evaluate the proposed use-case of post-OCR correction, we simulate the digitization pipeline. First, we synthetically create a PDF from each corpus in the following manner:

- We open the text file in a text editor. (We use `gedit` with `Sanskrit2003` font.)
- We next trigger the operating system's (Ubuntu 18.04) print dialogue (`Ctrl+P`)
- We use the '*Print to File*' option and save the file as PDF.

A PDF file created in this manner is the best-case scenario for OCR engines, as it contains no noise. Now, we run both the OCR systems and obtain the OCR-ed versions of the text. As a result of this simulation, we can now realistically evaluate the detection and correction of errors introduced in the process of OCR.

### 3.2 Results

We use the following abbreviations for different versions of the corpora:

- **WS**: Wikisource
- **SD**: sanskritdocuments.org
- **GR**: GRETIL
- **GO**: Google Drive OCR on a synthetically generated PDF
- **TO**: Tesseract OCR on a synthetically generated PDF

We evaluate our Chandojñānam system as well as those by Rajagopalan (2020) and Neill (2022) on each version of the corpora. We evaluate the ability of these systems to identify the chanda of the verse in the presence of errors. Table 3 illustrates the result of this evaluation. Chandojñānam was able to identify the correct meter from the erroneous text in 98.2% of the cases, performing better than the systems by Rajagopalan (2020) (91.9%) and Neill (2022) (80.3%). Corpora, code and results of the experiments are made available with the source code.

[16] `https://sa.wikisource.org/s/1c5`
[17] `https://sanskritdocuments.org/doc_z_misc_major_works/meghanew.html`
[18] `http://gretil.sub.uni-goettingen.de/gretil/1_sanskr/5_poetry/2_kavya/kmeghdpu.htm`
[19] `https://sa.wikisource.org/s/7xr`
[20] `https://sa.wikisource.org/s/7up`
[21] `https://sa.wikisource.org/s/7xq`

| | | Meghadūta | | | | | Śāntavilāsa | | | Rāmarakṣā | | | Rājendrakarṇapūra | | | Total |
|---|---|---|---|---|---|---|---|---|---|---|---|---|---|---|---|---|
| | | SD | GR | WS | GO | TO | WS | GO | TO | WS | GO | TO | WS | GO | TO | |
| Number of Verses | | 117 | 111 | 123 | 123 | 123 | 36 | 36 | 36 | 39 | 39 | 39 | 72 | 72 | 72 | 1038 |
| Unique Chanda | | 1 | 1 | 1 | 1 | 1 | 12 | 12 | 12 | 9 | 9 | 9 | 4 | 4 | 4 | 17 |
| Erroneous Verses | | 20 | 79 | 2 | 31 | 77 | 13 | 16 | 31 | 1 | 4 | 13 | 12 | 26 | 71 | 396 |
| Correct Meters Identified | Neill (2022) | 20 | 79 | 2 | 30 | 66 | 11 | 13 | 14 | 0 | 2 | 9 | 12 | 24 | 36 | 318 (80.3%) |
| | Rajagopalan (2020) | 19 | 79 | 2 | 30 | 75 | 12 | 15 | 24 | 1 | 2 | 9 | 12 | 26 | 58 | 364 (91.9%) |
| | Chandojñānam | 20 | 79 | 2 | 31 | 77 | 13 | 16 | 29 | 1 | 3 | 9 | 12 | 26 | 71 | 389 (98.2%) |

Table 3: Error tolerance of meter identification systems. (Versions are WS: Wikisource, GO: Google OCR, TO: Tesseract OCR, SD: sanskritdocuments.org, GR: GRETIL.) Chandojñānam is able to detect correct chanda from erroneous verses 98.2% of the times.

### 3.3 Error Analysis

We now analyze the errors in a more detailed manner.

Despite the Wikisource version being prepared through manual correction of OCR, Chandojñānam was still able to detect 2 errors from Meghadūta. The errors are described below:

- Line: कालक्षेपं ककुभसुरभौ पर्वते पर्वैते ते (Pāda 3, Śloka 1.23)
- Suggestion: [[['का', 'ल', 'क्षे', 'पं'], ['क', 'कु', 'भ', 'सु', 'र', 'भौ'], ['प', 'र्व', 'ते'], ['प', 'r(र्वै)[L]', 'ते'], ['ते']]]
- Description: The error is due to the incorrect word पर्वैते. It is likely that this was due to an oversight by the curator. It can be seen that the system correctly points to the location where a change is required.
- Line: साभिज्ञानप्रहितकुशलैस्ततद्वचोभिर्ममापि (Pāda 3, Śloka 2.53)
- Suggestion: [[['सा', 'भि', 'ज्ञा', 'न', 'प्र', 'हि', 'त', 'कु', 'श', 'लै', 'd(स्त)', 'त', 'द्व', 'चो', 'भि', 'र्म', 'मा', 'पि']]]
- Description: The error is due to an extra letter present in this line, where an extra त appears in the sandhi of words कुशलैः and तद्वत्, resulting in कुशलैस्ततद्वत् instead of कुशलैस्तद्वत्. The system is able to identify the error, and point out that a syllable needs to be deleted. However, we can see that the system points to an incorrect syllable स्त to be deleted. This can be explained from the fact that both स्त and त are laghu letters, and deletion of either letter results in the correct metrical signature. So, it is impossible for a meter identification based system to correctly say which character is to be deleted without any notion of semantic consideration. Such type of semantic error correction is out-of-scope as of yet.

Majority of the 'errors' in GRETIL stem from the text being written with sandhis being split. Despite not being linguistic errors, from the point of view of Chandaḥśāstra, these indeed are errors as they may change the number of syllables in a pāda and subsequently result in the change or loss of the meter.

It is important to remember that the errors detected are only the errors reported by the Chandojñānam system. Errors that do not result in the breaking of metrical pattern are impossible to be corrected by metrical analysis. It is interesting to note that the fuzzy matching performs better for Google OCR than Tesseract OCR. It can be explained as follows. As the Google OCR system performs better and makes fewer errors per line, the possible deviations are fewer. This results in less false positives. On the contrary, if the OCR system makes numerous errors, the *lg-signature* is farther away from the actual *lg-signature* and, therefore, 'the closest match' of the erroneous *lg-signature* might also be erroneous.

Finally, it can be seen that although the meter identification can be a useful tool for detection of errors in the Sanskrit text, there are several other factors that affect the error rate of such a system.

## 4 Conclusions

We have presented a Sanskrit meter identification tool that adds many user-friendly features and focuses on tolerance towards erroneous text as well as correction of such text. The features such as meter identification from images are useful for Sanskrit enthusiasts without much programming background. Bulk analysis of text files is a useful aid for digitization of Sanskrit texts using the methodology of post-OCR manual correction.

Future directions for the tool include addition of mātrā meters, improvement to meter detection and correction algorithm with the possible use of syntactic and semantic aspects, adding more extensive support for Indian languages that use similar rules of prosody.

The system is available online at `https://sanskrit.iitk.ac.in/jnanasangraha/chanda/` and the source code is available at `https://github.com/hrishikeshrt/chanda/`.